\documentclass[runningheads]{llncs}
\usepackage[english]{babel}
\usepackage{epsfig} 
\usepackage{amssymb}
\usepackage{amsfonts}
\usepackage{amsmath}
\usepackage{graphicx}

\def\A{\mathcal{A}}
\def\B{\mathcal{B}}

\def\O{\mathcal{O}}
\def\P{\mathcal{P}}

\def\q#1{\mathtt{#1}}
\def\dq#1{\mathtt{#1}}
\def\h{\q{h}}
\def\1{\q{1}}
\def\0{\q{0}}
\def\m{\q{\#}}
\def\j{\q{\_}}
\def\t{\q{@}}
\newcommand{\llabel}{\mathit{label}}

\title{A unifying framework for seed sensitivity and its application
  to subset seeds}
\subtitle{(Extended abstract)}
\titlerunning{A unifying framework for seed sensitivity}
\author{Gregory Kucherov\inst{1}
\and 
Laurent No{\'e}\inst{1}
\and 
Mikhail Roytberg\inst{2}
}
\institute{
INRIA/LORIA, 615, rue du Jardin Botanique, B.P. 101, 54602,
Villers-l\`es-Nancy, France, \email{\{Gregory.Kucherov,Laurent.Noe\}@loria.fr}
\and
Institute of Mathematical Problems in Biology, Pushchino, Moscow
Region, 142290, Russia, \email{roytberg@impb.psn.ru}
}
\authorrunning{G.~Kucherov, L.~No{\'e}, M.~Roytberg}

\date{{\normalsize{\today}}}

\begin{document}
\maketitle
\begin{abstract}
We propose a general approach to compute the seed sensitivity, that can
be applied to different definitions of seeds. It treats
separately three components of the seed sensitivity
problem --  a set of target alignments, an associated probability
distribution, and a seed model -- that are specified by
distinct finite automata. The approach is then applied to a new concept of
{\em subset seeds} for which we propose an efficient automaton
construction. Experimental results confirm that sensitive subset seeds
can be efficiently designed using our approach, and can then be used
in similarity search producing better results than ordinary spaced
seeds. 
  \end{abstract}%
  \section{Introduction}
In the framework of pattern
matching and similarity search in biological sequences, seeds specify
a class of short sequence motif which, if shared by two sequences, are
assumed to witness a potential similarity. 
Spaced seeds have been introduced several years ago
\cite{BurkhardtKarkkainen03,PatternHunter02} and have been shown to
improve significantly the efficiency of the search. One of the key
problems associated with spaced seeds is a 
precise estimation of the sensitivity of the associated search method.
This is important for comparing seeds and for choosing most
appropriate seeds for a sequence comparison problem to solve.

The problem of seed sensitivity depends on several
components. First, it depends on the {\em seed model} specifying the
class of allowed seeds and the way that seeds match ({\em hit}) potential
alignments. In the basic case, seeds are specified by
binary words of certain length ({\em span}), possibly with a constraint on
the number of 1's ({\em weight}). However, different extensions of this
basic seed model have been proposed in the literature, such as
multi-seed (or multi-hit) strategies
\cite{GBLAST97,BLAT02,PatternHunter02}, seed families
\cite{PatternHunter04,BuhlerRECOMB04,YangWangChenEtAlBIBE04,KucherovNoeRoytbergTCBB05,XuBrownLiMaCPM04,BrownTCBB05},
seeds over non-binary alphabets
\cite{ChenSungGI03,NoeKucherovBMC04}, vector seeds
\cite{BrejovaBrownVinarWABI03,BrownTCBB05}.

The second parameter is the class of {\em target alignments} that are
alignment fragments that one
aims to detect. Usually, 
these are {\em gapless} alignments of a given length. 
Gapless alignments are easy to model, in
the simplest case they are represented by binary sequences in the
match/mismatch alphabet. This representation has been adopted by many
authors
\cite{PatternHunter02,KeichLiMaTromp02,BrejovaBrownVinarJBCB04,ChoiZhang04,BuhlerKeichSunRECOMB03,CZZ04}.
The binary representation, however, cannot distinguish between different types
of matches and mismatches, and is clearly insufficient in the case
of protein sequences. In \cite{BrejovaBrownVinarWABI03,BrownTCBB05},
an alignment is represented by a sequence of real numbers that are
{\em scores} of matches or mismatches at corresponding positions. A
related, but yet different approach is suggested in
\cite{NoeKucherovBMC04}, where DNA alignments are represented by
sequences on the ternary alphabet of
match/transition/transversion. Finally, another generalization of
simple binary sequences was considered in
\cite{KucherovNoePontyBIBE04}, where alignments are required to be
{\em homogeneous}, i.e. to contain no sub-alignment with a score larger
than the entire alignment. 

The third necessary ingredient for seed sensitivity estimation is the
probability distribution on the set of target alignments. Again,
in the simplest case, alignment sequences are assumed to obey a
Bernoulli model \cite{PatternHunter02,ChoiZhang04}. In more general
settings, Markov or Hidden Markov models are considered
\cite{BuhlerKeichSunRECOMB03,BrejovaBrownVinarJBCB04}. A different way of defining
probabilities on binary alignments has been taken in
\cite{KucherovNoePontyBIBE04}: all homogeneous alignments of a given
length are considered equiprobable.

Several algorithms for computing the seed sensitivity for different
frameworks have been proposed in the above-mentioned
papers. All of them, however, use a common dynamic programming (DP)
approach, first brought up in \cite{KeichLiMaTromp02}.

In the present paper, we propose a general approach to computing the seed
sensitivity. This approach subsumes the cases
considered in the above-mentioned papers, and allows to deal with new
combinations of the three seed sensitivity parameters. 
%
The underlying idea of our approach is to specify each of the three
components -- the seed, the set of target alignments, and the
probability distribution -- by a separate finite automaton.

A deterministic finite automaton (DFA) that recognizes all
alignments matched by given seeds was already used in
\cite{BuhlerKeichSunRECOMB03} for the case of ordinary spaced seeds. 
In this paper, we assume that the set of target alignments is also
specified by a DFA and, more importantly, that the probabilistic model
is specified by a {\em probability transducer} -- a probability-generating
finite automaton equivalent to HMM with respect to the class of generated
probability distributions.


We show that once these three automata are set, the seed sensitivity
can be computed by a unique general algorithm. This algorithm reduces
the problem to a computation of the total weight over all paths in an
acyclic graph corresponding to the automaton resulting from the
product of the three automata. This computation
can be done by a well-known dynamic
programming algorithm \cite{AhoHopcroftUllman74,FinkRoytberg93} with the time
complexity proportional to the number of transitions of the
resulting automaton. Interestingly, all
above-mentioned seed sensitivity algorithms considered by different
authors can be reformulated as instances of this general algorithm.


In the second part of this work, we study a new concept of 
{\em subset seeds} -- an extension of spaced seeds that allows to deal
with a non-binary alignment alphabet and, on the other hand, still
allows an efficient hashing method to locate seeds. For this definition
of seeds, we define a DFA with a number of states independent of the
size of the alignment alphabet. Reduced to the case of ordinary spaced
seeds, this DFA construction gives the same worst-case number of
states as the Aho-Corasick DFA used in
\cite{BuhlerKeichSunRECOMB03}. Moreover, our DFA has always no
more states than the DFA of \cite{BuhlerKeichSunRECOMB03}, and
has substantially less states on average. 

Together with the general approach proposed in the first part, our DFA
gives an efficient algorithm for computing the sensitivity of subset
seeds, for different classes of target alignments and different
probability transducers. In the experimental part of this work, we confirm
this by running an implementation of our algorithm in order to design
efficient subset seeds for different probabilistic models, trained on
real genomic data. We also show experimentally that designed subset seeds
allow to find more significant alignments than ordinary spaced seeds
of equivalent selectivity.%
  \section{General Framework}\label{section:Framework}
Estimating the seed sensitivity amounts to compute the probability for
a random word (target alignment), drawn according to a given
probabilistic model, to belong to a given language, namely the
language of all alignments matched by a given seed (or a set of seeds).
\subsection{Target Alignments}
Target alignments are represented by words over an alignment alphabet
$\A$. In the simplest case, considered most often, the alphabet is
binary and expresses a match or a mismatch occurring at each alignment
column. However, it could be useful to consider larger alphabets,
such as the ternary alphabet of match/transition/transversion for the
case of DNA (see 
\cite{NoeKucherovBMC04}). The importance of this extension is even
more evident for the protein case (\cite{BrownTCBB05}), where
different types of amino acid pairs are generally distinguished. 

Usually, the set of target alignments is a finite set. In the case
considered most often
\cite{PatternHunter02,KeichLiMaTromp02,BrejovaBrownVinarJBCB04,ChoiZhang04,BuhlerKeichSunRECOMB03,CZZ04},
target alignments are all words of a given length $n$. This set is 
trivially a regular language that can be specified by a deterministic automaton
with $(n+1)$ states. 
However, more complex definitions of target alignments have been
considered (see e.g. \cite{KucherovNoePontyBIBE04}) that aim to capture more
adequately properties of biologically relevant alignments.
In general, we assume that the set of target alignments is a finite regular
language $L_T\in\A^*$ 
and thus can be represented by an acyclic DFA 
$T = <Q_T,q_T^0,q_T^F,\A,\psi_T>$.
\subsection{Probability Assignment}\label{subsection:proba-assign}
Once an alignment language $L_T$ has been set, we have to define a
probability distribution on the words of $L_T$. We do this 
using probability transducers.

A probability transducer is a finite automaton without final states in
which each transition outputs a {\em probability}. 
\begin{definition}
  \label{definition:probabilitytransducer}
  A {\em probability transducer} $G$ over an alphabet $\A$ is a 4-tuple
  $<Q_G,q_G^0,\A, \rho_G>$, where $Q_G$ is a finite set of states, $q_G^0\in Q_G$ is
  an initial state, and $\rho_G: Q_G\times \A\times Q_G\rightarrow [0,1]$ is
  a real-valued probability function such that \newline
  \mbox{$\forall q\in Q_G, \sum_{q'\in Q_G, a \in \A} \rho_G(q,a,q') = 1$}.
\end{definition}
A {\em transition} of $G$ is a triplet $e=<q,a,q'>$ 
such that $\rho(q,a,q')>0$. Letter $a$ is called the {\em label} of
$e$ and denoted $\llabel(e)$. A probability transducer $G$ is {\em deterministic}
if for each $q\in Q_G$ and each $a\in\A$, there is at most one
transition $<q,a,q'>$. 
For each path $P = ( e_1, ..., e_n )$ in $G$, we define its 
{\em label} to be the word
$\llabel(P) = \llabel(e_1)... \llabel(e_n)$, and the associated
probability to be the product $\rho(P) = \prod_{i=1}^n \rho_G(e_i)$.
A path is {\em initial}, if its start state is the initial state
$q_G^0$ of the transducer $G$.
\begin{definition}
  \label{definition:weightword}
The {\em probability} of a word $w\in \A^*$ according to a probability
transducer $G = <Q_G, q^0_G, \A,\rho_G>$, denoted $\P_G(w)$, is the
sum of probabilities of all initial paths in $G$ with the label
$w$. $\P_G(w)=0$ if no such path exists. 
The probability $\P_G(L)$ of a finite language $L\subseteq \A^*$ 
according a probability transducer $G$ is defined by $\P_G(L) =
\sum_{w\in L} \P_G(w)$.
\end{definition}

Note that for any $n$ and for $L=A^n$ (all words of length $n$), $\P_G(L) =1$. 

Probability transducers can express common probability distributions on
words (alignments). 
Bernoulli sequences with independent probabilities of each symbol
\cite{PatternHunter02,ChoiZhang04,CZZ04} 
can be specified with deterministic one-state probability transducers. 
In Markov sequences of order $k$
\cite{BuhlerKeichSunRECOMB03,BuhlerRECOMB04}, the probability of each
symbol depends on $k$ previous symbols. They can therefore be
specified by a deterministic probability transducer with at most
$|\A|^{k}$ states.

A Hidden Markov model (HMM) \cite{BrejovaBrownVinarJBCB04}
corresponds, in general, to a non-deterministic probability
transducer. The states of this transducer correspond to the (hidden)
states of the HMM, plus possibly an additional initial state. Inversely, for
each probability transducer, one can construct an HMM generating the
same probability distribution on words. Therefore, non-deterministic
probability transducers and HMMs are equivalent with respect to the
class of generated probability distributions. The proofs are
straightforward and are omitted due to space limitations. 

\subsection{Seed automata and seed sensitivity}
\label{subsection:seed-autom}

Since the advent of spaced seeds
\cite{BurkhardtKarkkainen03,PatternHunter02}, different extensions of
this idea have been proposed in the literature (see Introduction). For
all of them, the set of possible alignment fragments matched by a seed
(or by a set of seeds) is a finite set, and therefore the set of matched
alignments is a regular language. 
For the original spaced seed model, this observation was used by Buhler
et al.~\cite{BuhlerKeichSunRECOMB03} who proposed an algorithm for
computing the seed sensitivity based on a DFA defining the language of
alignments matched by the seed. 
In this paper, we extend this approach to a general one that allows a
uniform computation of seed sensitivity for a wide class of settings
including different probability distributions on target alignments, as
well as different seed definitions. 

Consider a seed (or a set of seeds) $\pi$ under a given seed model. 
We assume that the set of alignments $L_\pi$ matched 
by $\pi$ is a regular language recognized by a DFA
$S_\pi=<Q_S,q_S^0,Q_S^F,\A,\psi_S>$. Consider a finite set $L_T$ of
target alignments and a probability transducer $G$. 
Under this assumptions, the
sensitivity of $\pi$ is defined as the conditional probability 
\begin{equation}
\label{cond-proba}
\frac{\P_G(L_T \cap L_\pi)}{\P_G(L_T)}. 
\end{equation}
An automaton recognizing $L=L_T \cap L_\pi$ can be obtained 
as the product of automata $T$ and $S_\pi$ recognizing $L_T$ and
$L_\pi$ respectively. Let $K = <Q_K,q_K^0,Q_K^F,\A,$ $\psi_K>$ be this
automaton. 
We now consider the product $W$ of 
$K$ and $G$, denoted $K \times G$, defined as follows.
\begin{definition}
Given a DFA 
$K = <Q_K,q_K^0,Q_K^F,\A,\psi_K>$ and a probability transducer $G=<Q_G, q^0_G, \A,\rho_G>$,
the product of $K$ and $G$ is the {\em probability-weighted automaton}
 $W =<Q_W,q_W^0,Q_W^F,\A,\rho_W>$ (for short, {\em PW-automaton}) such that
$Q_W   =  Q_K \times Q_G$,  $q_W^0 =  (q_K^0, q_G^0)$, $q_W^F =  \{ (q_K,q_G) | q_K \in Q_K^F\}$,
$\rho_W((q_K, q_G),a,(q'_K,q'_G))  = \rho_G(q_G,a, q'_G)$ if
$\psi_K(q_K,a) = q'_K$, and $0$ otherwise.
\end{definition}
$W$ can be viewed as a non-deterministic probability transducer with
final states. $\rho_W((q_K,q_G),a,(q'_K,q'_G))$ is the 
{\em probability} of the  $<(q_K,q_G),a, (q'_K,q'_G)>$ transition. 
A path in $W$ is called {\em full} if it goes from the initial to a final
state. 
\begin{lemma}
\label{lemma:LangWeight}
Let $G$ be a probability transducer.
Let $L$ be a finite language and $K$ be a deterministic automaton
recognizing $L$. Let $W = G \times K$. 
The probability $\P_G(L)$ 
is equal to sum of probabilities of all full paths in $W$. 
\end{lemma}
\begin{proof}
Since $K$ is a deterministic automaton, each word $w\in L$ corresponds
to a single accepting path in $K$ and the paths in $G$ labeled $w$ (see
Definition~\ref{definition:probabilitytransducer}) are in one-to-one
correspondence with the full path in $W$ accepting $w$. 
By definition, $\P_G(w)$ is equal to the
sum of probabilities of all paths in $G$ labeled $w$. Each such path
corresponds to a unique path in $W$, with the same probability. Therefore,
the probability of $w$ is the sum of probabilities of corresponding paths in
$W$. Each such path is a full path, and paths for distinct words $w$
are disjoint. The lemma follows.
\end{proof}

\subsection{Computing Seed Sensitivity}
\label{subsection:algo}
Lemma \ref{lemma:LangWeight} reduces the computation of seed
sensitivity to a computation of the sum of probabilities of paths in a
PW-automaton. 

\begin{lemma}
\label{lemma:DP}
Consider an alignment alphabet $\A$, a finite set $L_T \subseteq \A^*$
of target alignments, and a set $L_{\pi} \subseteq \A^*$ of all
alignments matched by a given seed $\pi$. 
Let $K = <Q_K, q_t^0, Q_K^F, \A, \psi_Q>$ 
be an acyclic DFA recognizing the language $L=L_T \cap L_\pi$. Let
further $G = <Q_G, q_G^0, \A, \rho >$ be a probability transducer
defining a probability distribution on the set $L_T$. Then $\P_G(L)$
can be computed in time  $\O(|Q_G|^2\cdot|Q_K|\cdot|\A|)$ and space $\O(|Q_G|\cdot|Q_K|)$.
\end{lemma}

\begin{proof} By Lemma~\ref{lemma:LangWeight}, the probability
of $L$ with respect to $G$
can be computed as the sum of probabilities of all full paths in $W$. Since
$K$ is an acyclic automaton, so is $W$. Therefore, the sum of probabilities
of all full paths in $W$ leading to final states $q^F_W$ can  be
computed by a classical DP algorithm \cite{AhoHopcroftUllman74}
applied to acyclic directed graphs 
(\cite{FinkRoytberg93} presents a survey of application of this
technique to different bioinformatic  problems). 
The time complexity of the algorithm is proportional to the number of 
transitions in $W$. $W$ has $|Q_G|\cdot |Q_K|$ states, and for each
letter of $\A$, each state has at most $|Q_G|$ outgoing
transitions. The bounds follow. 
\end{proof}

Lemma~\ref{lemma:DP} provides a general approach to compute the seed
sensitivity. To apply the approach, one has to define three automata:
\begin{itemize}
\item a deterministic acyclic DFA $T$ specifying a set of
  target alignments over an alphabet $\A$ (e.g. all words of a given
  length, possibly verifying some additional properties), 
\item a (generally non-deterministic) probability transducer $G$ specifying
  a probability distribution on target alignments (e.g. Bernoulli
  model, Markov sequence of order $k$, HMM), 
\item a deterministic DFA $S_\pi$ specifying the seed model via a set
  of matched alignments.
\end{itemize}
As soon as these three automata are defined, Lemma~\ref{lemma:DP} can
be used to compute probabilities $\P_G(L_T\cap L_\pi)$ and $\P_G(L_T)$
in order to estimate the seed sensitivity according to equation (\ref{cond-proba}).

Note that if the probability transducer $G$ is deterministic (as it is the
case for Bernoulli models or Markov sequences), then the time
complexity is $\O(|Q_G|\cdot|Q_K|\cdot|\A|)$. 
In general, the
complexity of the algorithm can be improved by reducing the involved
automata. Buhler et al. \cite{BuhlerKeichSunRECOMB03} introduced the
idea of using the Aho-Corasick automaton~\cite{AhoCorasick74} as the seed automaton
$S_\pi$ for a spaced seed. The authors
of \cite{BuhlerKeichSunRECOMB03} considered all binary alignments of a
fixed length $n$ distributed according to a Markov model of order $k$. In
this setting, the obtained complexity was $\O(w2^{s-w}2^k n)$, where
$s$ and $w$ are seed's span and weight respectively. Given that the
size of the Aho-Corasick automaton is $\O(w2^{s-w})$, this complexity
is automatically implied by
Lemma~\ref{lemma:DP}, as the size of the
probability transducer is $\O(2^k)$, and that of the target alignment
automaton is $\O(n)$. Compared to \cite{BuhlerKeichSunRECOMB03},
our approach explicitly distinguishes the descriptions of matched
alignments and their probabilities, which allows us to automatically
extend the algorithm to more general cases. 

Note that the idea of using the Aho-Corasick automaton can be applied
to more general seed models than individual spaced seeds (e.g. to
multiple spaced seeds, as pointed out in
\cite{BuhlerKeichSunRECOMB03}). In fact, all currently proposed seed
models can be described by a finite set of matched alignment
fragments, for which the Aho-Corasick automaton can be constructed. We
will use this remark in later sections.

The sensitivity of a spaced seed with respect to an HMM-specified
probability distribution over binary target alignments of a given length $n$
was studied by Brejova et al. \cite{BrejovaBrownVinarJBCB04}. 
The DP algorithm of \cite{BrejovaBrownVinarJBCB04} has a lot in common
with the algorithm implied by
Lemma~\ref{lemma:DP}. In 
particular, the states of the algorithm of
\cite{BrejovaBrownVinarJBCB04} are triples 
$<w, q, m>$, where $w$ is a prefix of the seed $\pi$, $q$ is a state
of the HMM, and $m\in [0..n]$.  
The states therefore correspond to the construction implied by
Lemma~\ref{lemma:DP}. However, the authors
of~\cite{BrejovaBrownVinarJBCB04} do not consider any automata,
which does not allow to optimize the preprocessing step (counterpart
of the automaton construction) and, on the other hand, does not allow to extend the algorithm to more
general seed models and/or different sets of target alignments. 

A key to an efficient solution of the sensitivity problem remains the
definition of the seed. It should be expressive enough to be able to
take into account properties of biological sequences. On the other
hand, it should be simple enough to be able to locate seeds fast and
to get an efficient algorithm for computing seed
sensitivity. According to the approach presented in this section, the
latter is directly related to the size of a DFA specifying the seed.%
  \section{Subset seeds}
\label{section:SubsetSeeds}
 
\subsection{Definition}
\label{subsection:SubsetSeeds}
Ordinary spaced seeds use the simplest possible binary ``match-mismatch''
alignment model that allows an efficient implementation by hashing all
occurring combinations of matching positions. 
A powerful generalization of spaced seeds, called {\em vector seeds},
has been introduced in \cite{BrejovaBrownVinarWABI03}. 
Vector seeds allow one to use an arbitrary alignment alphabet
and, on the other hand, provide a flexible definition of a hit based
on a cooperative contribution of seed positions. 
A much higher expressiveness of vector seeds lead to more complicated 
algorithms and, in particular, prevents the application of direct
hashing methods at the seed location stage. 

In this section, we consider {\em subset seeds} that have an 
intermediate expressiveness between spaced and vector seeds. It allows
an arbitrary alignment alphabet and, on the other hand, still allows
using a direct hashing for locating seed, which maps each
string to a unique entry of the hash table. 
We also propose a construction of a seed automaton for subset seeds,
different from the Aho-Corasick automaton. The automaton has
$\O(w 2^{s-w})$ states {\em regardless of the size of the alignment
  alphabet}, where $s$ and $w$ are respectively the span of the seed
and the number of ``must-match'' positions. From the general algorithmic
framework presented in the previous section (Lemma~\ref{lemma:DP}),
this implies that the seed sensitivity can be computed for subset
seeds with same complexity as for ordinary spaced seeds. 
Note also that for the binary alignment
alphabet, this bound is the same as the one implied by the Aho-Corasick
automaton. However, for larger alphabets, the Aho-Corasick construction
leads to $\O(w|\A|^{s-w})$ states. In the experimental part of this
paper (section~\ref{subsection:aut-size}) we will show that even for
the binary alphabet, our automaton construction yields a smaller
number of states in practice. 

Consider an alignment alphabet $\A$. We always assume that $\A$
contains a symbol $\1$, interpreted as ``match''. 
A {\em subset seed} is defined as a word over a {\em seed alphabet}
$\B$, such that
\begin{itemize}
\item letters of $\B$ denote subsets of the alignment
alphabet $\A$ containing $\1$ ($\B \subseteq \{\1\}\cup 2^{\A}$), 
\item  $\B$ contains a letter $\m$ that denotes subset $\{\1\}$,
\item a subset seed $b_1 b_2\ldots b_m\in\B^m$ matches an alignment fragment
$a_1 a_2\ldots a_m\in\A^m$ if $\forall i \in [1..m]$, $a_i \in b_i$. 
\end{itemize}
The {\em $\m$-weight} of a subset seed $\pi$ is the number of $\m$
in $\pi$ and the {\em span} of $\pi$ is its length.

\begin{example}
\label{example-subset}
\cite{NoeKucherovBMC04} considered
the alignment alphabet $\A = \{\1,\h,\0\}$ representing respectively a
match, a transition mismatch, or a transversion mismatch in a DNA
sequence alignment. The seed alphabet is $\B = \{\m,\t,\j\}$ denoting
respectively subsets $\{\1\}$, $\{\1,\h\}$, and $\{\1,\h,\0\}$. 
Thus, seed $\pi = \dq{\#@\_\#}$ matches alignment $s = \dq{10h1h1101}$
at positions $4$ and $6$. The span of $\pi$ is $4$, and the
$\m$-weight of $\pi$ is 2. 
\end{example}
Note that unlike the weight of
ordinary spaced seeds, the $\m$-weight cannot serve as a measure of
seed selectivity. In the above example, symbol $\t$ should be assigned
weight $0.5$, so that the weight of $\pi$ is equal to $2.5$
(see~\cite{NoeKucherovBMC04}).

\subsection{Subset Seed Automaton}
Let us fix an alignment alphabet $\A$, a seed alphabet $\B$, and a
seed $\pi=\pi_1\pi_2\ldots \pi_{m} \in \B^*$ of span $m$ and $\m$-weight
$w$. Let $R_{\pi}$ be the set of all non-$\m$ positions in
$\pi$, $|R_{\pi}|= r = m - w$.
We now define an automaton 
$S_{\pi} = <Q, q_0, Q_f, \A,\psi : Q\times\A\to Q>$ 
that recognizes the set of all alignments matched by $\pi$. 

The states $Q$ of $S_{\pi}$ are pairs $<X,t>$ such that 
$X \subseteq R_{\pi}, t \in [ 0,\ldots,m ]$, with the following
invariant condition. 
Suppose that $S_{\pi}$ has read a prefix $s_1\ldots s_p$ of an
alignment $s$ and has come to a state $<X,t>$. 
Then $t$ is the length of the longest suffix of $s_1\ldots s_p$ of the
form $\1^i$, $i\leq m$, 
and $X$ contains all positions $x_i\in R_\pi$ 
such that prefix $\pi_1 \cdots \pi_{x_i}$ of $\pi$ matches a suffix of
$s_{1}\cdots s_{p-t}$. 
\def\PiMotif#1#2{
  \put(#1,#2){
     \put(0,0){\large$\mbox{\tt \#}$}
     \put(1,0){\large$\mbox{\tt  @}$}
     \put(2,0){\large$\mbox{\tt \#}$}
     \put(3,0){\large$\mbox{\tt \_}$}
     \put(4,0){\large$\mbox{\tt \#}$}
     \put(5,0){\large$\mbox{\tt \#}$}
     \put(6,0){\large$\mbox{\tt \_}$}
     \put(7,0){\large$\mbox{\tt \#}$}
     \put(8,0){\large$\mbox{\tt \#}$}
     \put(9,0){\large$\mbox{\tt \#}$}
  }
}
\def\StrMotif#1#2{
  \put(#1,#2){
     \put(0,0){\large$\mbox{\tt 1}$}
     \put(1,0){\large$\mbox{\tt 1}$}
     \put(2,0){\large$\mbox{\tt 1}$}
     \put(3,0){\large$\mbox{\tt h}$}
     \put(4,0){\large$\mbox{\tt 1}$}
     \put(5,0){\large$\mbox{\tt 0}$}
     \put(6,0){\large$\mbox{\tt 1}$}
     \put(7,0){\large$\mbox{\tt 1}$}
     \put(8,0){\large$\mbox{\tt h}$}
     \put(9,0){\large$\mbox{\tt 1}$}
     \put(10,0){\large$\mbox{\tt 1}$} 
     \put(11,0){\large$\mbox{\tt ...}$}
  }
}
\def\PisevenPref#1#2{
  \put(#1,#2){
     \put(0,0){\large$\mbox{\tt \#}$}
     \put(1,0){\large$\mbox{\tt  @}$}
     \put(2,0){\large$\mbox{\tt \#}$}
     \put(3,0){\large$\mbox{\tt \_}$}
     \put(4,0){\large$\mbox{\tt \#}$}
     \put(5,0){\large$\mbox{\tt \#}$}
     \put(6,0){\large$\mbox{\tt \_}$}
  }
}
\def\PifourPref#1#2{
  \put(#1,#2){
     \put(0,0){\large$\mbox{\tt \#}$}
     \put(1,0){\large$\mbox{\tt  @}$}
     \put(2,0){\large$\mbox{\tt \#}$}
     \put(3,0){\large$\mbox{\tt \_}$}
  }
}
\def\PitwoPref#1#2{
  \put(#1,#2){
     \put(0,0){\large$\mbox{\tt \#}$}
     \put(1,0){\large$\mbox{\tt  @}$}
  }
}
\vspace{-0.5cm}
\begin{figure*}[htb]\center
  \begin{picture}(100,40)(0,0)\noindent\centering\setlength{\unitlength}{6pt}
    \put(-18, 4){$(a)$}
    \put(-15, 4){\large$\pi = $}\PiMotif{-11}{ 4}
    \put(-18, 1){$(b)$}
    \put(-15, 1){\large$ s = $}\StrMotif{-11}{ 1}
    \put(10,4){$(c)$}
    \put(22.5, 7){$s_9$}
    \put(23.5, 6.5){\line(0,-1){0.5}}
    \put(24.2, 6.5){\line(0,-1){0.5}}
    \put(25.8, 6.5){\line(0,-1){0.5}}
    \put(24.2, 6.5){\line(1, 0){1.6}}
    \put(24.7, 7){$t$}
    \StrMotif{15}{5}
    \put(13, 3.25){$\pi_{1..7}=$}\PisevenPref{17}{3} 
    \put(16, 1.75){$\pi_{1..4}=$}\PifourPref{20}{1.5}
    \put(18, 0.25){$\pi_{1..2}=$}\PitwoPref{22}{0}
  \end{picture}
\caption{\label{figure:Statedef} Illustration to Example~\ref{example-state}}
\end{figure*}
\vspace{-0.5cm}
\begin{example}
\label{example-state}
In the framework of Example~\ref{example-subset}, 
consider a seed $\pi$ and an alignment prefix $s$ 
of length $p = 11$ given on Figure~\ref{figure:Statedef}(a) and~\ref{figure:Statedef}(b) respectively. The length $t$ of the last run of
$\1$'s of $s$ is $2$. The last mismatch position of $s$ is $s_9 = \h$.
The set $R_\pi$ of non-$\m$ positions of $\pi$ is $\{2,4,7\}$ and 
$\pi$ has 3 prefixes ending at positions of $R_\pi$ (Figure~\ref{figure:Statedef}(c)). 
Prefixes $\pi_{1..2}$ and $\pi_{1..7}$ do match suffixes of 
$s_1 s_2\ldots s_9$, and prefix $\pi_{1..4}$ does not.
Thus, the state of the automaton after reading $s_1 s_2\ldots s_{11}$ is
$<\{2,7\}, 2>$.\\
\end{example}

The initial state $q_0$ of $S_{\pi}$ is the state $ <\emptyset, 0>$.
The final states $Q_f$ of $S_{\pi}$ are all states $q = <X,t>$, where
$max\{X\} + t = m$. All final states are merged into one state.


The transition function $\psi(q,a)$ is defined as follows: 
If $q$ is a final state, then $\forall a \in \A$, $\psi(q,a) = q$. 
If $q = <X,t>$ is a non-final state, then
\begin{itemize}
\item if $a = \1$ then $\psi(q,a) = <X,t+1>$, 
\item otherwise $\psi(q,a) =  <X_{U} \cup X_{V},0>$ with
  \begin{itemize}
  \item $X_{U} = \{ x           | x \leq t + 1 \mathrm{~and~} a \in \pi_{x} \}$
  \item $X_{V} = \{ x + t + 1   | x \in X  \mathrm{~and~} a \in \pi_{x+t+1}\}$
  \end{itemize}
\end{itemize}

\begin{lemma}\label{lemma:Gpi}
The automaton $S_{\pi}$ accepts the set of all alignments matched by $\pi$.
\end{lemma}
\begin{proof}
It can be verified by induction that the invariant condition on the states
$<X,t>\in Q$ is preserved by the transition function $\psi$. 
The final states verify $max\{X\} + t = m$, which implies
that $\pi$ matches a suffix of $s_1\ldots s_p$. 
\end{proof}

\begin{lemma}\label{lemma:GnbStates}
The number of states of the automaton $S_{\pi}$ is no more than $(w+1)2^{r}$.
\end{lemma}
\begin{proof}
Assume that $R_\pi=\{x_1,x_2,\ldots,x_r\}$ 
and $x_1 < x_2 \cdots < x_r$. Let $Q_i$ be the set of non-final states
$<X,t>$ with $max\{X\} = x_i$, $i\in [1..r]$.
For states $q = <X,t> \in Q_i$ there are $2^{i-1}$
possible values of $X$ and $m-x_i$ possible values of 
$t$,
as $max\{X\}+t\leq m-1$. Thus, $|Q_i| \;\leq\; 2^{i-1} (m-x_i) \leq  2^{i-1} (m-i)$ and $\sum_{i=1}^{r}   |Q_i|   \;\leq\; \sum_{i=1}^{r} 2^{i-1} (m-i)  =  (m-r+1) 2^r - m - 1$.
Besides states $Q_i$, $Q$ contains $m$ states $<\emptyset,t>$
($t\in [0..m-1]$) and one final state. 
Thus, $|Q| \leq (m-r+1) 2^r = (w+1) 2^r$.
\end{proof}

Note that if $\pi$ starts with $\m$, which is always the case for
ordinary spaced seeds, then $X_i \geq i+1$, $i\in[1..r]$, and previous
bound rewrites to $2^{i-1} (m-i-1)$. 
This results in the same number of states $w 2^r$ as for
the Aho-Corasick automaton \cite{BuhlerKeichSunRECOMB03}.
The construction of automaton $S_\pi$ is optimal, in the sense that no
two states can be merged in general.
%
A straightforward generation of the transition table of the automaton
$S_\pi$ can be performed in time
$\O(r \cdot w \cdot 2^{r} \cdot |\A| )$. A more complicated algorithm
allows one to reduce the bound  to $\O(w \cdot 2^{r} \cdot
|\A|)$. 
In the next section, we demonstrate experimentally that on average,
our construction yields a very compact automaton, close to the minimal
one. Together with the general approach of
section~\ref{section:Framework}, this provides a fast algorithm
for computing the sensitivity of subset seeds and, in turn, allows to
perform an efficient design of spaced seeds well-adapted to the
similarity search problem under interest. 
  \section{Experiments}
\label{section:experiments}
Several types of experiments have been performed to test the practical
applicability of the
results of sections~\ref{section:Framework},\ref{section:SubsetSeeds}. 
We focused on DNA similarity search, and set the
alignment alphabet $\A$ to $\{\1,\h,\0\}$ (match, transition, 
transversion). For subset seeds, the seed alphabet $\B$ was set to $
\{\m,\t,\j\}$,  where 
$\m = \{\1\}, \t = \{\1,\h\}, \j = \{\1,\h,\0\}$ (see
Example~\ref{example-subset}). 
The weight of a subset seed is computed by assigning weights $1$,
$0.5$ and $0$ to symbols $\m$, $\t$ and $\j$ respectively. 
\subsection{Size of the automaton}
\label{subsection:aut-size}
%
We compared the size of the automaton $S_\pi$ defined in
section~\ref{section:SubsetSeeds} and the
Aho-Corasick automaton~\cite{AhoCorasick74}, both for ordinary spaced
seeds (binary seed alphabet) and for subset seeds (ternary seed alphabet). The Aho-Corasick
automaton for spaced seeds was constructed as defined in
\cite{BuhlerKeichSunRECOMB03}. For subset seeds, a straightforward
generalization was considered: the Aho-Corasick construction was
applied to the set of alignment fragments matched by the seed. 

Tables~\ref{table:Automaton1}(a) and \ref{table:Automaton1}(b) present
the results for spaced seeds and subset seeds respectively. For each
seed weight $w$, we computed the average number of states ($avg.$ $s.$)
of the Aho-Corasick automaton and our automaton $S_\pi$, and reported
the corresponding ratio ($\delta$) with respect to the average number of states of the
minimized automaton. The average was computed over all seeds of span
up to $w+8$ for spaced seeds and all seeds of span up to $w+5$ with two $\t$'s
 for subset seeds.
\vspace{-0.5cm}
\begin{table*}[htb]%
      \begin{center}%
        {\scriptsize%
        \begin{tabular}{c|cc|cc|cc}%
          {\bf Spaced} &
          \multicolumn{2}{|c}{Aho-Corasick}&
          \multicolumn{2}{|c}{$S_\pi$}&
          \multicolumn{1}{|c}{Minimized}\\
          $w$       &
          $avg.$ $s.$  &
          $\delta$      &
          $avg.$ $s.$  &
          $\delta$      &
          $avg.$ $s.$  \\
          \hline
          \hline
          9  & 345.94 & 3.06 & 146.28 & 1.29 & 113.21 \\
          10 & 380.90 & 3.16 & 155.11 & 1.29 & 120.61 \\
          11 & 415.37 & 3.25 & 163.81 & 1.28 & 127.62 \\
          12 & 449.47 & 3.33 & 172.38 & 1.28 & 134.91 \\
          13 & 483.27 & 3.41 & 180.89 & 1.28 & 141.84 \\
        \end{tabular}%
        ~~
        \begin{tabular}{c|cc|cc|cc}
          {\bf Subset} &
          \multicolumn{2}{|c}{Aho-Corasick}&
          \multicolumn{2}{|c}{$S_\pi$}&
          \multicolumn{1}{|c}{Minimized}\\
          $w$       &
          $avg.$ $s.$  &
          $\delta$      &
          $avg.$ $s.$  &
          $\delta$      &
          $avg.$ $s.$  \\
          \hline
          \hline
          9  & 1900.65 & 15.97  & 167.63 & 1.41 & 119.00\\
          10 & 2103.99 & 16.50  & 177.92 & 1.40 & 127.49\\
          11 & 2306.32 & 16.96  & 188.05 & 1.38 & 135.95\\
          12 & 2507.85 & 17.42  & 198.12 & 1.38 & 144.00\\
          13 & 2709.01 & 17.78  & 208.10 & 1.37 & 152.29\\
        \end{tabular}
        }
        ~~~~~~~~~(a)~~~~~~~~~~~~~~~~~~~~~~~~~~~~~~~~~~~~~~~~~~~~~~~~~~~~~~~~~~~~~~~~~~~~~~~~~~~~~~~~~~~~~~(b)
        \caption{\it \label{table:Automaton1} Comparison of the
          average number
          of states of Aho-Corasick automaton, automaton $S_\pi$ of
          section~\ref{section:SubsetSeeds} and minimized automaton}%
      \end{center}%
    \end{table*}%
\vspace{-0.8cm}
Interestingly, our automaton turns out to be more compact
than the Aho-Corasick automaton
not only on non-binary alphabets (which was expected), but also on the
binary alphabet (cf Table~\ref{table:Automaton1}(a)). Note that for a
given seed, one can define a 
surjective mapping from the states of the Aho-Corasick
automaton onto the states of our automaton. This implies that our
automaton has {\em always} no more states than the Aho-Corasick
automaton. 
\subsection{Seed Design}

In this part, we considered several probability transducers
to design spaced or subset seeds. The target alignments included all
alignments of length $64$ on alphabet $\{\1,\h,\0\}$. Four
probability transducers have been studied 
(analogous to those introduced in \cite{BrejovaBrownVinarCPM03}): 
\begin{itemize}
\item $B$: Bernoulli model
\item $DT1$: deterministic probability transducer specifying
  probabilities of $\{\1,\h,\0\}$ at each codon position 
(extension of the $M^{(3)}$ model of~\cite{BrejovaBrownVinarCPM03} to the
three-letter alphabet), 
\item $DT2$: deterministic probability transducer specifying
  probabilities of each of the 27 codon instances $\{\1,\h,\0\}^3$
  (extension of the $M^{(8)}$ model of~\cite{BrejovaBrownVinarCPM03}),
\item $NT$: non-deterministic probability transducer combining four
  copies of $DT2$ specifying four distinct codon conservation levels
  (called HMM model in \cite{BrejovaBrownVinarCPM03}).
\end{itemize}
Models $DT1$, $DT2$ and $NT$ have been trained on alignments resulting
from a pairwise comparison of $40$ bacteria genomes. 
\begin{table*}[htb]%
  \begin{center}%
    \begin{tabular}{c|lc|lc}
      {$w$} &
      {~~spaced seeds} &
      {~~Sens.} & 
      {~~subset seeds, two $\t$} &
      {~~Sens.} \\
      \hline
      \hline
      9  &~\mbox{\tt \#\#\#\_\_\_\#\_\#\_\#\#\_\#\#}~&       0.4183
      &~\mbox{\tt \#\#\#\_\#\_\_\#@\#\_@\#\#}~&           0.4443\\
      10 &~\mbox{\tt \#\#\_\#\#\_\_\_\#\#\_\#\_\#\#\#}~&     0.2876
      &~\mbox{\tt \#\#\#\_@\#\_@\#\_\#\_\#\#\#}~&         0.3077\\
      11 &~\mbox{\tt \#\#\#\_\#\#\#\_\#\_\_\#\_\#\#\#}~&     0.1906
      &~\mbox{\tt \#\#@\#\_\_\#\#\_\#\_\#\_@\#\#\#}~&     0.2056\\
      12 &~\mbox{\tt \#\#\#\_\#\_\#\#\_\#\_\_\#\#\_\#\#\#}~& 0.1375
      &~\mbox{\tt \#\#@\#\_\#\_\#\#\_\_\#@\_\#\#\#\#}~&   0.1481\\
    \end{tabular}
    \caption{\it \label{table:Bernoulli} Best seeds and their sensitivity for probability transducer $B$}
  \end{center}%
\end{table*}%
\begin{table*}[htb]%
    \begin{center}
             {
	       \begin{tabular}{c|lc|lc}
		 {$w$} &
		 {~~spaced seeds} &
		 {~~Sens.} & 
		 {~~subset seeds, two $\t$} &
		 {~~Sens.} \\
		 \hline 
		 \hline
		 9 & \mbox{\tt \#\#\#\_\_\_\#\#\_\#\#\_\#\# }~& 0.4350
		 & \mbox{\tt \#\#@\_\_\_\#\#\_\#\#\_\#\#@ }~& 0.4456\\
		 10 & \mbox{\tt \#\#\_\#\#\_\_\_\_\#\#\_\#\#\_\#\# }~& 0.3106
		 & \mbox{\tt \#\#\_\#\#\_\_\_@\#\#\_\#\#@\# }~& 0.3173\\
		 11 & \mbox{\tt \#\#\_\#\#\_\_\_\_\#\#\_\#\#\_\#\#\# }~& 0.2126
		 & \mbox{\tt \#\#@\#@\_\#\#\_\#\#\_\_\#\#\# }~& 0.2173\\
		 12 & \mbox{\tt \#\#\_\#\#\_\_\_\_\#\#\_\#\#\_\#\#\#\# }~& 0.1418
		 & \mbox{\tt \#\#\_@\#\#\#\_\_\#\#\_\#\#@\#\# }~& 0.1477\\
	     \end{tabular}}%
	     \caption{\it \label{table:M3} Best seeds and their sensitivity for probability transducer $DT1$}
    \end{center}%
\end{table*}%
\begin{table*}[htb]%
    \begin{center}
               \begin{tabular}{c|lc|lc}
                 {$w$} &
                 {~~spaced seeds} &
                 {~~Sens.} & 
                 {~~subset seeds, two $\t$} &
                 {~~Sens.}\\
                 \hline 
                 \hline
		 9  &~\mbox{\tt \#\_\#\#\_\_\_\_\#\#\_\#\#\_\#\#}~&         0.5121
		 &~\mbox{\tt \#\_\#@\_\#\#\_@\_\_\#\#\_\#\#}~&           0.5323\\
		 10 &~\mbox{\tt \#\#\_\#\#\_\#\#\_\_\_\_\#\#\_\#\#}~&       0.3847
		 &~\mbox{\tt \#\#\_@\#\_\#\#\_\_@\_\#\#\_\#\#}~&         0.4011\\
		 11 &~\mbox{\tt \#\#\_\#\#\_\_\#\_\#\_\_\_\#\_\#\#\_\#\#}~& 0.2813
		 &~\mbox{\tt \#\#\_\#\#\_@\#\_\#\_\_\_\#\_\#@\_\#\#}~&   0.2931\\
		 12 &~\mbox{\tt \#\#\_\#\#\_\#\#\_\#\_\_\_\#\_\#\#\_\#\#}~& 0.1972
		 &~\mbox{\tt \#\#\_\#\#\_\#@\_\#\#\_@\_\_\#\#\_\#\#}~&   0.2047\\
               \end{tabular}
	     \caption{\it \label{table:M14} Best seeds and their sensitivity for probability transducer $DT2$} 
    \end{center}%
\end{table*}%
\begin{table*}[htb]%
  \begin{center}
             \begin{tabular}{c|lc|lc}
               {$w$} &
               {~~spaced seeds} &
               {~~Sens.} & 
               {~~subset seeds, two $\t$} &
               {~~Sens.} \\              
               \hline 
               \hline
	       9  &~\mbox{\tt \#\#\_\#\#\_\#\#\_\_\_\_\#\#\_\#}  & 0.5253
	       &~\mbox{\tt \#\#\_@@\_\#\#\_\_\_\_\#\#\_\#\#} & 0.5420\\   
	       10 &~\mbox{\tt \#\#\_\#\#\_\_\_\_\#\#\_\#\#\_\#\#} & 0.4123        
	       &~\mbox{\tt \#\#\_\#\#\_\_\_\_\#\#\_@@\_\#\#\_\#} & 0.4190\\        
	       11 &~\mbox{\tt \#\#\_\#\#\_\_\_\_\#\#\_\#\#\_\#\#\_\#} & 0.3112        
	       &~\mbox{\tt \#\#\_\#\#\_\_\_\_\#\#\_@@\_\#\#\_\#\#} & 0.3219\\       
	       12 &~\mbox{\tt \#\#\_\#\#\_\_\_\_\#\#\_\#\#\_\#\#\_\#\#} & 0.2349        
	       &~\mbox{\tt \#\#\_\#\#\_\_\_\_\#\#\_@@\_\#\#\_\#\#\_\#} & 0.2412\\
             \end{tabular}
	   \caption{\it \label{table:HMM} Best seeds and their sensitivity for probability transducer $NT$} 
  \end{center}%
\end{table*}%
For each of the four probability transducers, we computed the best
seed of weight $w$ ($w=9,10,11,12$) among two categories: ordinary
spaced seeds of weight $w$ and subset seeds of weight $w$ with two
$\t$. Ordinary spaced seeds were enumerated exhaustively up to a given
span, and for each seed, the sensitivity was computed using the
algorithmic approach of section~\ref{section:Framework} and the seed
automaton construction of section~\ref{section:SubsetSeeds}. Each such
computation took between 10 and 500ms on a Pentium IV 2.4GHz computer
depending on the seed weight/span and the model used. In each
experiment, the most sensitive seed found has been kept. The results
are presented in Tables~\ref{table:Bernoulli}-\ref{table:HMM}.

In all cases, subset seeds yield a better sensitivity than ordinary
spaced seeds. The sensitivity increment varies up to 0.04 which is a
notable increase. As shown in \cite{NoeKucherovBMC04}, the gain in using subset
seeds increases substantially when the transition probability is
greater than the transversion probability, which is very often the case
in related genomes.
  \section{Discussion}
\label{section:discussion}

We introduced a general framework for computing the seed sensitivity
for various similarity search settings.  The approach 
can be seen as a generalization of methods of
\cite{BuhlerKeichSunRECOMB03,BrejovaBrownVinarJBCB04}  
in that it allows to obtain algorithms with the same worst-case
complexity bounds as those proposed in these papers, but also allows
to obtain efficient algorithms for new formulations of the seed
sensitivity problem. This versatility is achieved by distinguishing
and treating separately the three ingredients of the seed sensitivity
problem: a set of target alignments, an associated probability
distributions, and a seed model.

We then studied a new concept of {\em subset seeds} which represents an interesting
compromise between the efficiency of spaced seeds and the flexibility
of vector seeds. For this type of seeds, we defined an automaton with
$\O(w2^r)$ states regardless of the size of the alignment alphabet,
and showed that its transition table can be constructed in time
$\O(w2^r|\A|)$. Projected to the case of spaced seeds, this
construction gives the same worst-case bound as the Aho-Corasick
automaton of \cite{BuhlerKeichSunRECOMB03}, but results in a smaller
number of states in practice. Different experiments we have done
confirm the practical efficiency of the whole method, both at the level
of computing sensitivity for designing good seeds, as well as using
those seeds for DNA similarity search. 

As far as the future work is concerned, it would be interesting to
study the design of efficient spaced seeds for protein sequence
search (see \cite{BrownTCBB05}), as well as to combine spaced seeds
with other techniques such as 
seed families
\cite{PatternHunter04,BuhlerRECOMB04,KucherovNoeRoytbergTCBB05} or the
group hit criterion \cite{NoeKucherovBMC04}. 

\paragraph{Acknowledgments}
%
G.~Kucherov and L.~No\'e have been supported
by the {\em ACI IMPBio} of the French Ministry of Research. 
M.~Roytberg has been also supported by the Russian Foundation for
Basic Research (projects 03-04-49469, 02-07-90412) and by grants
from the RF Ministry of Industry, Science and Technology (20/2002,
5/2003) and NWO (Netherlands Science Foundation).

  \bibliographystyle{splncs}
  \bibliography{paper}
\end{document}